\documentclass{article}
\usepackage[utf8]{inputenc}
\usepackage[margin=1in]{geometry}
\usepackage{amsmath}
\usepackage{graphicx}
\usepackage{hyperref}
\usepackage{csquotes}
\usepackage{natbib}
\usepackage{parskip}
\title{On the simulation of the Hawkes process via Lambert-W functions}
\author{Martin Magris\footnote{Tampere University, Faculty of Information Technology and Communication Sciences, P.O. Box 541, FI-33101 Tampere, Finland. Email: martin.magris@tuni.fi}}

\date{May 2019}

\begin{document}

\maketitle
\begin{abstract}
    Several methods have been developed for the simulation of the Hawkes process. The oldest approach is the inverse sampling transform (ITS) suggested in \citep{ozaki1979maximum}, but rapidly abandoned in favor of more efficient alternatives. This manuscript shows that the ITS approach can be conveniently discussed in terms of Lambert-W functions. An optimized and efficient implementation suggests that this approach is computationally more performing than more recent alternatives available for the simulation of the Hawkes process.
\end{abstract}

\section{Introduction}

Forty years passed since the very first method for the simulation of the so-called Hawkes process passed \citep{ozaki1979maximum}. However, the method developed in \citep{ozaki1979maximum} has been abandoned very soon in favor of the more efficient method of \citep{lewis1979simulation}, applied to the Hawkes process in \citep{ogata1981lewis}, only two years later. In the following decades, further improvements on the simulation of the Hawkes process and alternative methods have been explored, leaving the simple and intuitive methods of \citep{ozaki1979maximum} widely un-applied, outdated and obsolete. 

This manuscript wants to re-discover the method of \citep{ozaki1979maximum}, revising him under a new perspective allowed by computational availability of the present days. 
These pages I go back to the very first simulation algorithm of \citep{ozaki1979maximum}  providing and insight on an efficient inverse transform sampling (ITS) -based simulation. Given three alternative-but-not-competing approaches for the simulation of the Hawkes process, this discussion focuses on improving the efficiency of the oldest (and less-sophisticated) one. In particular, showing that its main drawback (which constitutes the inefficiency of the method) can be conveniently resolved by the use of Lambert-W functions. By doing so, the original simulation problem is thus reduced to a straightforward sequence of simple-functions evaluations.

This is not a proper scientific novelty, indeed is a re-interpretation of a very-old approach in a way that modern computing capabilities allow it to by practically applied much more efficiently by using a convenient framework. As I show later, the simulation method of \citep{ozaki1979maximum} solves in the independent evaluation of sequences of Lambert-W functions. If on one side this sound quite appealing, i.e. simulating means simple function evaluations, on the other hand, the Lambert-W function is quite complex, non-standard and of difficult implementation. I.e. the problem is shifted from the complexity of the simulation itself to the efficient computation of the Lambert-W function. However, if the latest task is properly addressed the revised method of \citep{ozaki1979maximum} can perhaps outperform most of the recently-developed and widespread alternatives.

\section{Alternative approaches to the simulation of the Hawkes' process}
In the past years, a number of articles discussed the problem of the simulation of the Hawkes process. There are three main-streams under which the simulation of the Hawkes process can be implemented. 
\begin{itemize}
\item As precisely described in Section \ref{sec:arrivaltimes}, the intensity function of the Hawkes process can be used to recover the conditional distribution of the inter-arrival times, leading to a straightforward simulation procedure. In particular, given a random variable $X$ and its CDF the standard way to simulate from $X$ is via \enquote{inverse transform sampling} (ITS). This is the most standard procedure for simulating from a random variable $X$: the method can be applied to the Hawkes process as well. The ITS-based simulation of the Hawkes process goes back to \citep{ozaki1979maximum}. Indeed, given the popularity of the ITS method, it's not surprising that \cite{ozaki1979maximum} is the earliest work on the simulation of the Hawkes process. Interestingly, no further improvements and works over the findings of \citep{ozaki1979maximum} have been lately suggested and published. \citep{ozaki1979maximum} algorithm is clearly inefficient since it involves a time-consuming numerical (likely Newton-Raphson) approximation for each event being simulated. Later on, the method of \citep{ogata1981lewis}  gained popularity as a standard simulation method, and the ITS method has been in practice abandoned.

    \item First of all, being the Hawkes process a generic non-homogeneous Poisson process the general procedure for the simulation of the non-homogeneous Poisson process applies. This is the so-called \enquote{Thinning algorithm}. The theoretical foundation of this procedure goes back to \citep{lewis1979simulation}, which provides a \textit{general procedure} for the simulation of a \textit{generic} non-homogeneous process. This result has first been used in the context of the Hawkes process by \citep{ogata1981lewis}, often referred to as \enquote{Ogata's modified thinning algorithm}. Later, this has been extended to the multivariate case too \citep[e.g.][]{ogata1998space}.
    
    \item \citep{hawkes1974cluster} provide an immigration-birth representation of the Hawkes process. In this context, the Hawkes process is viewed as a superimposition of a first process (immigrants) which generate a cascade of descendant events (births). By this reasoning is possible to simulate the Hawkes process in such a way that neither thinning nor ITS are involved. The representation in \cite{hawkes1974cluster} attains to the Hawkes process only and by this representation, we do not get a general simulation procedure for the non-homogeneous Poisson process as with the Thinning method. However, the simulation based on the immigrant-birth representation is attractive over the Ogata's thinning and definitely outperforms the ITS of \citep{ozaki1979maximum} (since no CDF inversion and not complex equations are involved). The literature on the simulation of the Hawkes process in the last years mostly evolved all around the immigrant-birth representation \citep[e.g.]{moller2005perfect, moller2006approximate}.  The recent algorithm of \citep{dassios2013exact} constitutes the endpoint of the literature advances: a fast and efficient method for the simulation of the Hawkes process with an exponential kernel based on the immigration-birth representation (no rejection-sampling and scales linear to the number of events drawn).
\end{itemize}

In discussion shows that a convenient and simple algebraic manipulation of the original equations of \citep{ozaki1979maximum} allow to review this leading work in terms of Lambert-W functions, and simulate the underlying Hawkes process accordingly. 

\section{Arrival times simulation via IST}\label{sec:arrivaltimes}
\subsection{CDF of the inter-arrival times}
Be $t_0,...,t_i,...,t_n$ the arrival times of $n+1$ events and let $d_0,...,d_i,...,d_n$ denote the durations, $d_i = t_{i+1}-t_i$. The underlying process for which $\left\lbrace t_i \right\rbrace_{i=0,...,n}$ represents a sample, is characterized by the conditional intensity function $\lambda\left( t | H_t, \theta \right)$ (or $\lambda_t$). The conditional intensity depends on the past history of the process (up to time $t$), $H_t = \left\lbrace  t_0,...,t_n \right\rbrace$ and a vector of parameters $\theta$.\\
The conditional intensity $\lambda$ uniquely defines the point process and can be used to recover the conditional CDF of the inter-event times (durations):
$$ F_t = Pr\left( t_{n+1} \leq t | H_t, \theta \right) = 1-e^{-\int_{t_n}^t \lambda_s ds}$$
where $t_n$ is the last event before $t$.
In other words, given the realizations $t_0,...,t_n$ which define $H_t$ and a parameter $\theta$, $F_t$  is the CDF characterizing the random arrival time $t_{n+1}$. $F_t$ is entirely and uniquely specified only by the intensity $\lambda$.\\
Here $F_t$ is defined in terms of the absolute timestamp $t$: a more convenient notation is to define the CDF in terms of $t-t_{n}$, i.e. time passed after $t_n$, duration $t-t_n$. With this simple re-parametrization $F_t$ turns into (omitting $H_t$ and $\theta$ from the notation):
$$ F_{\delta,t_{n}} = Pr\left( t_{n+1}-t_n \leq \delta \right) = 1-e^{-\int_{t_n}^{t_n+\delta} \lambda_s ds} $$
Note that since $\lambda$ is conditional on $H_t$ in general $F_{\delta,t_i} \neq  F_{\delta,t_j}$ for $i \neq j$, therefore at each $t_i$, the CDF describing the duration to the next event changes as $H_{t_{i-1}}$ updates to $H_t$, thus affecting $\lambda_t$.

\subsection{Inverse transform sampling - The Ozaki (1979) approach}\label{sec:its}
A standard approach for random number sampling is the so-called inverse transform sampling. Given the CDF $F_X$ and its inverse $F^{-1}_X$, we can generate a random draw $x$ form $X$ by:
$$ x = F_x^{-1}\left( u \right), \; \; u \in \left[0,1\right] $$
In general, given a uniform random variable $U$ over $\left[0,1\right]$, $X = F^{-1}_X\left( U \right)$.\\
For a complex CDF $F$ the inverse $F_X^{-1}$ may not be available in analytic form. A turnaround to apply the inverse transform sampling is to numerically solve (for $x$) the equation:
$$ F_X\left( x \right) = u$$
For any $u$, a solution $x^*$ is by construction a draw from $X$.\\
Going back to the introduction, this procedure can be exploited for simulating the arrival time $t_{n+1}$ given that the last event is $t_n$ and $H_t$ is known. The solution $\delta^*$, solving
$$ F_{\delta,t_n} - u = 0 $$
is a randomly sampled duration from $F_{\delta,t_n}$ of the time between $t_{n+1}$ (random at $t_n)$ and $t_{n}$ (observed in $t_n$ and included in $H_{t_n}$).\\

\subsection{The Ozaki algorithm}
The \citep{ozaki1979maximum} algorithm is a straightforward application of the ITS method over the CDF $F_X$ of the inter-arrival times of the Hawkes process. This distribution is explicitly addressed in \ref{sec:its}, where the mathematical discussion on the Lambert-W approach is carried out. The following resembles the original simulation approach:
\begin{itemize}
    \item[i] Given a starting point $t_0$ and a set of parameters $\theta$, compute $\lambda_0$.
    \item[ii] Generate a random uniform draw $u \in \left[0,\right]$, obtain $\delta ^*$ solving for $\delta$: $F_{\delta,0}-u = 0$.
    \item[iii]Set $t_1 = t_0 + \delta$.
    \item[iv] Iterate by repeating steps ii and iii: taring from $t_1$, obtain $t_2$ and so on.
\end{itemize}{}
The set $\left\lbrace t_0,t_1,...\right\rbrace$ is then a simulated path from the process characterized by the intensity $\lambda \left( t | H_t, \theta \right)$.
The major drawback of this simple and straightforward approach is the numerical solution required in (ii): in this form, the algorithm is clearly inefficient and there's no surprise that alternative methods have been developed.

\section{Arrival times simulation via Lambert-W functions}
\subsection{Lambert-W function}\label{sec:Lambert}
For a complez number $z$ consider the function $f\left(z \right) = z e^z$. The inverse function $f^{-1}\left( ze^z \right)$ is the so-called Lambert-W function:
$$ z = f^{-1}\left( ze^z \right) = W\left(ze^z \right)$$
Given a general problem in the form $ze^z = w$, its solution ($z^*$) is therefore $z^* = W\left(w \right)$.\\
Consider a non-linear equation in the form:
\begin{equation}\label{eq:general_nonlin_eq}
ae^x+bx+c = 0
\end{equation}
a solution for $x$ can be easily obtained by use of the $W$ function.\\
Set $y = bx+c$, then eq. \eqref{eq:general_nonlin_eq} rewrites as $ae^{\frac{y-c}{b}}+y = 0$.
After some emelentary algebra, this rearanges as:
\begin{equation}\label{eq:general_nonlin_solution}
    -\frac{y}{b}e^{-\frac{y}{b}} = \frac{a}{b}e^{-\frac{c}{b}}
\end{equation}
Equation \eqref{eq:general_nonlin_solution} is in the form $ze^z =w$, whose solution is $W(w)$: it rewrites as $bx+c = -bW\left( \frac{a}{b}e^{-\frac{c}{b}} \right)$. 
Then, the general solution of eq. \eqref{eq:general_nonlin_eq} is:
\begin{equation}\label{eq:general_nonlin_solution_x}
    x = -W\left( d \right) - \frac{c}{b} \; \; \; \text{with} \; \; \; d = \frac{a}{b}e^{-\frac{c}{b}}
\end{equation}

\subsection{Inverse transform sampling - The \enquote{Lambert} approach}
Consider the uni-variate self-exciting counting process $N_t$ whose intensity is given by:
\begin{equation}\label{eq:general_intensity}
  \lambda\left(t \right) = \mu + \int_{-\infty}^t g\left( t -u \right) dN\left(d \right)  = \mu + \sum_{t_k < t} g(t-t_k)  
\end{equation}

and consider the response function $g(t) = \alpha e^{-bt}$. Given the events $t_0,...,t_k$, the conditional CDF $F_{t_k,\delta}$ is given by:
$$F_{t_k,\delta} = 1-e^{-\int_{t_k}^{t_k+\delta}  \mu + \alpha \sum_{i = 1}^k e^{-\beta(t-t_i)}}$$

The intergal in the exponential solves to:
\begin{flalign*}
\int_{t_k}^{t_k+\delta}  \mu + \alpha \sum_{i = 1}^k e^{-\beta(t-t_i)} &= \mu\delta + \alpha \sum_{i = 1}^k \int_{t_k}^{t_k+\delta} e^{-\beta(t-t_i)} 
 = \mu \delta - \frac{\alpha}{\beta}  \left[e^{-\beta \delta} -1 \right] S_k
\end{flalign*}
where $S_k$ is the sum-of-exponents over all the $k$ time instances, $S_k =\sum_{i = 1}^k  e^{-\beta\left( t_k - t_i \right)}$.\\
According to section \ref{sec:its}, to simulate a sample duration $t_{n+1}-t_i$ for the time to the next event after $t_n$, one needs to solve ($u \in \left[ 0,1 \right]$):
\begin{equation}\label{eq:zero_problem}
F_{t_k,\delta} - u = 0    
\end{equation}
By rewriting eq. \eqref{eq:zero_problem} by use of the above integration, one immediately rewrites eq. \eqref{eq:zero_problem}:
$$\alpha S_k e^{x} + \mu x +  \left[ -\beta \log\left( 1-u\right) -\alpha S_k \right] = 0$$
where $x$ replaces $-\beta \delta $ by $x$.
This is a non-linear equation in the form \eqref{eq:general_nonlin_eq}, with $x = -\beta \delta$, $A = \alpha S_k$, $B = \mu$ and  $C = -\beta \log\left( 1-u\right) -\alpha S_k$.\\
By equation \eqref{eq:general_nonlin_solution_x}, $-\beta \delta = W\left(d \right) -\frac{C}{B}$, therefore:

\begin{equation}\label{eq:LambertW_to_simplify}
\begin{split}
\delta &= \frac{1}{\beta}\left[ W\left(d\right) - \frac{C}{B}\right]\\ 
d &=\frac{A}{B}e^{-\frac{C}{B}}, \; \; \; A = \alpha S_k, \; \; \; B = \mu, \; \; \; C = -\beta \log \left( 1-u \right) -\alpha S_k
\end{split}
\end{equation}
By some algebra on $d$ and $\frac{C}{B}$ one obtains the following convenient representations:
\begin{equation} \label{eq:thers_d_CB}
d = \frac{\alpha S_k}{\mu} \left( 1-u\right)^{\frac{\beta}{\mu}}e^{\frac{\alpha S_k}{\mu}} \; \; \; \; \; \; \;
\frac{C}{B} = \log \left[  \left( 1-u\right)^{\frac{\beta}{\mu}} \right]  + \frac{\alpha S_k}{\mu} 
\end{equation}
Note if that $U\sim Unif\left[ 0,1 \right]$, then $1-U$ distributes as a uniform distribution on $\left[0,1 \right]$ as well.
Equations \eqref{eq:LambertW_to_simplify} and \eqref{eq:thers_d_CB} lead to the final set of simplified equations, from which the solution of \eqref{eq:zero_problem} is immediately recovered:

\begin{subequations}
\label{eq:LWfinal}
 \begin{align}
  \delta &= \frac{1}{\beta} \left[ W\left(d\right) -\log B - A\right] \label{eq:LWfinal_a} \\ 
  d &= ABe^A, \; \; \; A = \frac{\alpha}{\mu}S_k, \; \; \; B = u^{\frac{\beta}{\mu}} \label{eq:LWfinal_b}
 \end{align}
\end{subequations}

\subsection{The Lambert algorithm}
    \begin{itemize}
        \item[i] Generate a random draw $r$ from an exponential distribution with parameter $\mu$.\footnote{This is a standard procedure in the context of the Hawkes process. Since there's no history to condition the intensity on, the first draw is generated assuming $\lambda = \mu$, corresponding to the intensity of an exponential distribution.} Set $t_0 = r.$
        \item[ii] Compute the quantities in \eqref{eq:LWfinal_b} and get $\delta$ from equation \eqref{eq:LWfinal_a}. Set $t_1 = t_0 + \delta$.
        \item[iii] Iterate [ii]. Assume $t_0,...,t_i$ are available (have been generated), compute the quantities in equation \eqref{eq:LWfinal_a}, compute $\delta$ from equation \eqref{eq:LWfinal_a} and set $t_{i+1} = t_i + \delta$.
    \end{itemize}
    
\section{Performance against competing algorithms}

As a benchmark for evaluating the simulation method here described (hereafter called Lambert) I implemented three relevant alternatives. (i) The inverse transform sampling as in \cite{ozaki1979maximum}, where the Lambert argument has not been discussed, and eq. \eqref{eq:zero_problem} is numerically solved. (ii) Ogata's thinning algorithm \citep{ogata1981lewis}, which is likely the most common choice for the simulation of the Hawkes process. (iii) \cite{dassios2013exact} procedure based on the cluster representation of the Hawkes process.

\begin{figure}[h!]
\centering
\includegraphics[scale=0.5,angle=0]{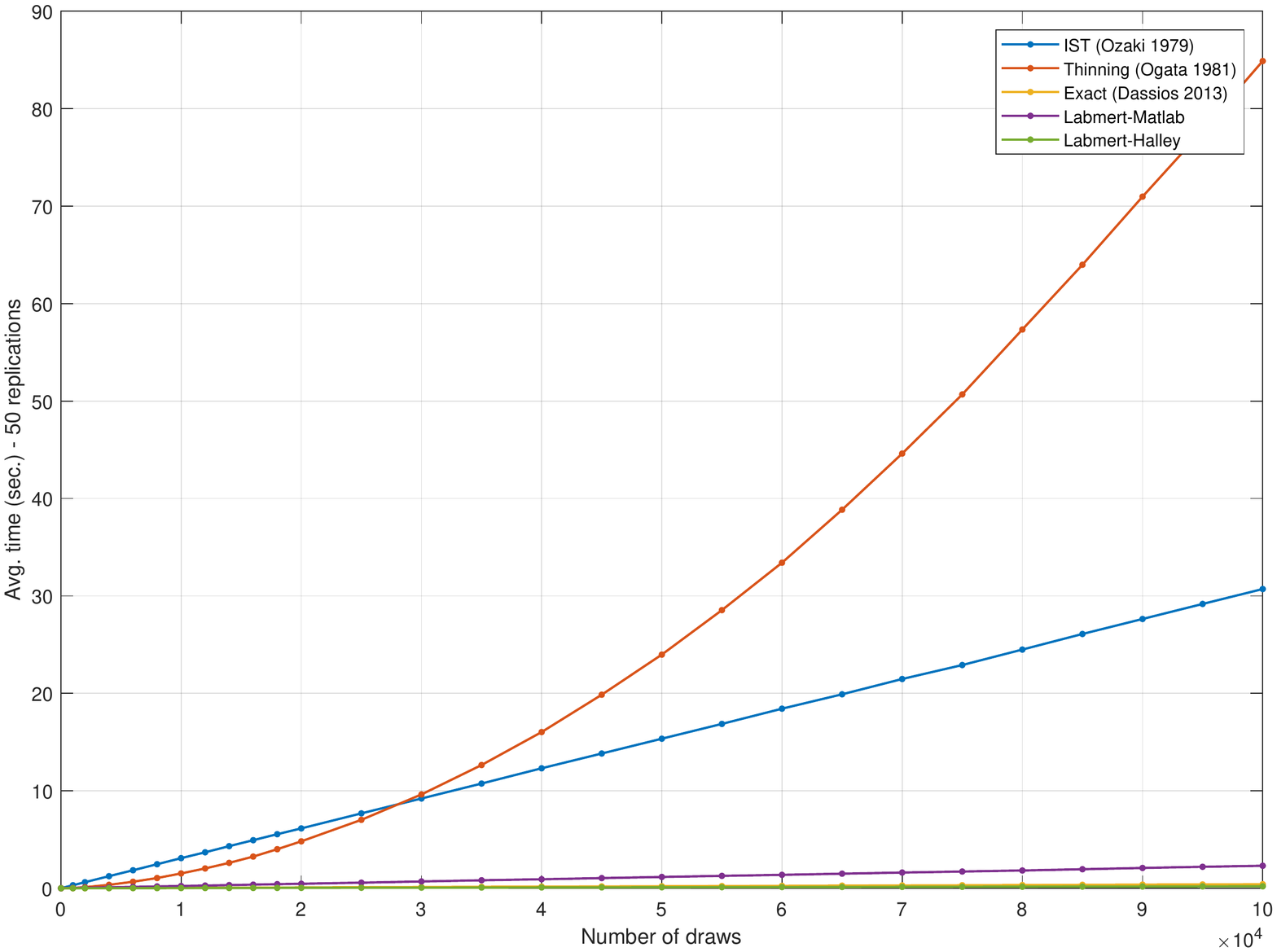}
\caption{Comparison between different algorithms for the simulation of the Hawkes process. Note: \enquote{Lambert-matalb} corresponds to the implementation with Matlab's \texttt{lambertw} function, \enquote{Matlab-Halley} is the implementation over my own Lambert-W function by using Halley's method.}
\label{fig:comparison}
\end{figure}

\textbf{Inverse sampling transform}.
The simulation algorithm discussed in \cite{ozaki1979maximum} exploits the standard simulation method based on the inverse sampling transform and is the closest algorithm to the Lambert method here introduced. However in \cite{ozaki1979maximum}  the exact solution of equation \eqref{eq:general_nonlin_eq} is not addressed. The zero of the transcendental equation \eqref{eq:general_nonlin_eq} is found numerically via Newton-Raphson method. This is the main drawback of the method ITS of \cite{ozaki1979maximum}: the solution is approximate and the root of eq. \eqref{eq:general_nonlin_eq} is found via a time-consuming numerical approximation  \citep[as][points out]{ogata1981lewis}. The Lambert simulation solves both the issues, providing exact solutions and avoiding any numerical approximation. Solutions of eq. \eqref{eq:general_nonlin_eq} are simply computed evaluating eq. \eqref{eq:general_nonlin_solution_x}.  As a consequence the gain in terms of efficiency is outstanding, see Fig. \ref{fig:comparison}.

\textbf{Thinning}.
A common technique for generating an in-homogeneous Poisson process is via thinning algorithm, first introduced in \cite{lewis1979simulation}. The intuitive idea is that of simulating a candidate point from a homogeneous process which is kept or removed probabilistically in such a way that the set of all the remaining points satisfy the time-varying intensity $\lambda_t$. A very similar approach is the so-called Ogata modified thinning algorithm \citep{ogata1981lewis}. While \citep{lewis1979simulation} require an almost sure upper bound $M$ for $\lambda_t$, given the non-increasing behaviour of the intensity \eqref{eq:general_intensity} (with $g\left(t\right)  = \alpha e^{-\beta t}$) in periods without arrivals, it is practically possible to identify an upper bound for $\lambda_t$ at any $t$. By recalling the left-continuity of $\lambda_t$ (predictability given $H_t$) and assuming that the intensity jump size of at every event is not greater than $\alpha$: $\lambda\left( t_i^+\right) \geq \lambda \left( t \right)$ with $t\in \left[ t_i,t_{i+1}\right]$. Therefore for any $t$ in any interval $\left[ t_i,t_{i+1}\right]$, $M_t = \lambda\left( t_i^+\right)$ is an upper bound for $\lambda_t$. By updating $M_t$ at every simulated event, one defines a piece-wise function bounding $\lambda_t$ at any $t$ (and $M_0 = \mu$). \\
The probabilistic pruning involved in the thinning algorithm constitutes its main drawback: a candidate point is kept (becomes an actual event of the process) with some probability, otherwise another random point is generated and so on, until the condition is randomly met. 
Therefore the algorithm does not provide a sample of length $n$ in a fixed amount of steps and a precise amount of time. Also, $M$ needs to be updated at every step (regardless or not if the random condition is met) by evaluating $\lambda_t$, which can be time-consuming, especially when the simulated process is long (many draws). Moreover, the algorithm indeed requires the simulation of two random numbers at each step, not just one and the execution time scales non-linearly with the number of draws.

\textbf{Exact simulation}.
Among the simulations algorithms based on the immigration-birth representation of the Hawkes process \citep{hawkes1974cluster}, the simulation procedure proposed in \cite{dassios2013exact} is the one chosen as a benchmark. Alternatives like \cite{moller2005perfect,moller2006approximate} may suffer from edge effects and are computationally more complex than \cite{dassios2013exact}. An advantage of \cite{dassios2013exact} over the thinning algorithm is the (i) absence of rejection sampling (no draws are discarded based on some probabilistic condition) and (ii) linear time-scaling of the number of draws. 

Fig. \ref{fig:comparison} provides a clear and immediate outlook of the performance of the four selected algorithms. Not too much to comment over the relative efficiency of the Lambert method against \cite{ozaki1979maximum} and \cite{ogata1981lewis}. 
The algorithm of \cite{dassios2013exact} can be outperformed by the Lambert algorithm depending on how efficiently the Lambert-W is implemented, Matlab's inbuilt \texttt{lambertw} function is not a basic implementation: the argument can be reals, vector, matrices of real and complex numbers (this negative one as well), whereas the algorithm only requires the evaluation of the Lambert-W function for positive real numbers. Therefore I wrote my own function which does exactly what is needed and nothing more, leading to an important gain in runtime. The custom function is based on Halley's method \citep[see e.g.][]{veberic2010having}. This is the difference between the lines marked as \enquote{Lambert-Matlab} and \enquote{Lambert-Halley} in Fig. \ref{fig:comparison}.

\section{Concluding remarks}
The Lambert method rediscovers the feasibility of the ITS for the Hawkes process, in such a way that nowadays's most standard approach (Ogata's thinning) is greatly outperformed by the Lambert method. This is quite interesting considering that the Ozaki method has been abandoned more 40 years ago, lacking for a clear direction for any improvement. The math here presented is very elementary and reduced to basic algebra:  the apparent complexity of the ITS involving the CDF inversion is overcome by proper use of Lambert-W functions as in eq. \eqref{eq:general_nonlin_solution_x}. 
Importantly, as the figure clearly shows, the performance of the Lambert algorithm depends on the efficiency of the Lambert-W function implementation, as well as the efficiency in the implementation of all the other methods. So far, no further improvements can be done on the codes here implemented, however, it's hard to argue that the plot does not clearly identify a \enquote{winning} algorithm.\\
Perhaps this is not enough to show that the Lambert approach outperforms the other ones on a general basis, but for sure it rediscovers the outdated method of \citep{ozaki1979maximum} finding him, based on nowadays' fast computers and excellent computing environments, capable of competing with the most recent alternatives, when re-arranged in terms of Lambert-W functions.

\bibliographystyle{apalike} 
\bibliography{BibFile} 

\end{document}